\documentclass[prl,twocolumn]{revtex4-1}

\usepackage{graphicx}  
\usepackage{dcolumn}  
\usepackage{subfigure}
\usepackage{amsmath}
\usepackage{natbib}

\newcommand{\EqKXe}{\left( \kappa_{0} \right)_{\mathrm{RbXe}}}
\newcommand{\EqKHe}{\left( \kappa_{0} \right)_{\mathrm{RbHe}}}

\newcommand{\He}{$^3$He}
\newcommand{\Xe}{$^{129}$Xe}
\newcommand{\KXe}{$\left( \kappa_{0} \right)_{\mathrm{RbXe}}$}
\newcommand{\KHe}{$\left( \kappa_{0} \right)_{\mathrm{RbHe}}$}
\newcommand{\XeD}{$\left[\mathrm{Xe}\right]$}

\begin{document}
\title{Collisional \He\ and \Xe\ frequency shifts in Rb--noble-gas mixtures}
\author{Z.\ L.\ Ma}
\author{ E.\ G.\ Sorte}
\author{B.\ Saam}
\email{saam@physics.utah.edu}
\affiliation{Department of Physics and Astronomy, University of Utah, 115 South 1400 East, Salt Lake City, Utah, 84112-0830, USA}
\begin{abstract}
The Fermi-contact interaction that characterizes collisional spin exchange of a noble gas with an alkali-metal vapor also gives rise to NMR and EPR frequency shifts of the noble-gas nucleus and the alkali-metal atom, respectively.  We have measured the enhancement factor $\kappa_{0}$ that characterizes these shifts for Rb-\Xe\ in the high-pressure limit to be 493$\pm$31, making use of the previously measured value of $\kappa_{0}$ for Rb-\He. This result allows accurate \Xe\ polarimetry with no need to reference a thermal-equilibrium NMR signal.
\end{abstract}
\maketitle
\noindent

The study of hyperpolarized noble gases generated by spin-exchange optical pumping (SEOP) \cite{Walker1997} continues to be vital and integral to recent work in many other fields, including condensed matter physics \cite{Morgan2008}, materials science \cite{Wang2009}, and medical imaging \cite{Yablonskiy2009,Driehuys2009}. A fundamental aspect of SEOP physics is the collisional Fermi-contact hyperfine interaction $\alpha {\bf K} \cdot {\bf S}$ between the noble-gas nuclear spin $\bf{K}$ and the alkali-metal electron spin $\bf{S}$, where $\alpha$ is the coupling strength. This interaction is not only responsible for spin-exchange hyperpolarization of noble gases, but it also gives rise to complementary shifts in both the NMR frequency of the noble gas and the EPR frequency of the alkali-metal vapor that are proportional to the electron and nuclear magnetizations, respectively \cite{Schaefer1989}.  These shifts provide insight into the nature of the interatomic potentials that ultimately determine spin-exchange rates for a given alkali-metal--noble-gas pair. If properly calibrated, the EPR shift also offers a simple and robust means to do noble-gas polarimetry in a typical low-field (few gauss) SEOP apparatus. The enhancement factor $\kappa_0$ that characterizes the frequency-shift calibration has been successfully measured for Rb-$^3$He to about 2\% \cite{Romalis1998,Barton1994} but until now was known to only about 50\% for Rb-\Xe\ \cite{Schaefer1989}. The Rb-\Xe\ measurement presents several method-dependent challenges, among them the fact that, unlike helium, high densities of xenon are difficult to polarize by SEOP. Indeed, the current lean-xenon flow-through method for generating large quantities of highly polarized \Xe\ \cite{Ruset2006,Driehuys1996} would benefit greatly from a more precise measurement of $\kappa_0$ for Rb-\Xe. In this work, we make consecutive measurements of the NMR shifts of both \He\ and \Xe\ at 2~T in the same glass cell under steady-state SEOP conditions. In cells having relatively low Xe density ($\left[\mathrm{Xe}\right]\lesssim$~10~Torr at 20~$^{\circ}$C) we use the ratio of these shifts to deduce a much more precise temperature-independent value,
\begin{equation}
\label{KappaValue}
\EqKXe = 493\pm 31,
\end{equation}

\noindent from the known value of \KHe.  In cells having  $\left[\mathrm{Xe}\right]$~approx.~ten~times~greater we observed an anomalous depression $\left(\approx 20\%\right)$ of the shift ratio at the highest temperatures.

\noindent

Averaged over many collisions, the interatomic hyperfine coupling results in a NMR frequency shift \cite{Schaefer1989}, 
\begin{equation}
\label{Shift}
\Delta\left|\nu_{X}\right|=-\frac{1}{h}\frac{\left|\mu_{K}\right|}{K}\frac{8\pi}{3}\mu_Bg_S\kappa_{XA}[A]\langle
S_z\rangle ,
\end{equation}

\noindent where $X$ is the noble gas species, $h$ is Planck's constant, $\mu_K$ is the nuclear magnetic moment, $\mu_B$ is the Bohr magneton, $g_S\approx 2$ is the Land$\mathrm{\acute{e}}$ factor, $\left[A\right]$ is the alkali-metal number density, and $\langle S_z\rangle$ is the volume-averaged expectation value of the \textit{z}-component of the alkali-metal electron spin (in units of $\hbar$). The dimensionless factor $\kappa_{XA}$ accounts for the enhancement of $\Delta\nu_X$ over the value it would have if the electron-spin magnetization were distributed continuously across a spherical sample.  This shift is analogous to the Knight shift first observed in metallic copper \cite{Townes1950}.  An equation complementary to Eq.~\eqref{Shift} yields the EPR shift $\Delta\nu_A$ of the alkali-metal electron in the presence of the nuclear magnetization (proportional to $\left[X\right]\langle K_z\rangle$) with enhancement factor $\kappa_{AX}$ \cite{Schaefer1989}.  In the limit of high gas densities \cite{Schaefer1989} that holds for all of this work, $\kappa_{XA}=\kappa_{AX}\equiv\kappa_0$ for both Rb-\He\ and Rb-\Xe; i.e., van der Waals molecules play a negligible role in determining the shifts.

\noindent

A direct measurement of \KXe\ using Eq.~\eqref{Shift} requires a measurement of the Rb magnetization (proportional to [Rb]$\langle S_z\rangle$). To avoid this and substantially simplify the experiment, we form the ratio
\begin{equation}
\label{ShiftRatio}
\EqKXe=\EqKHe\left(\frac{\gamma_{\rm He}}{\gamma_{\rm Xe}}\right)\left(\frac{2\Delta\nu_{\rm Xe}}{2\Delta\nu_{\rm He}}\right),
\end{equation}

\noindent where $\gamma_X$ are the noble-gas gyromagnetic ratios and $2\Delta\nu_{\rm X}$ are the shifts in the respective noble-gas NMR frequencies when the Rb vapor is exactly flipped from the low- to the high-energy Zeeman polarization state (LES and HES, respectively); $|{\rm [Rb]}\langle S_z\rangle |$ is presumed to remain constant under steady-state SEOP conditions. We note that ``LES" and ``HES" will be used strictly in reference to the Rb polarization state and never to that of the \He\ or \Xe. In this work, we measure directly the frequency-shift ratio in Eq.~\eqref{ShiftRatio}, averaging many measurements to reduce the statistical uncertainty. We then multiply by the previously measured $\EqKHe = 4.52 + 0.00934T$ \cite{Romalis1998} (presumed valid over our temperature range), where $T$ is the temperature in $^{\circ}$C, to deduce \KXe.

\noindent

Measurements were made on six \textit{d}=7~mm i.d.\ sealed uncoated Pyrex-glass spheres containing a few milligrams of naturally abundant Rb metal along with \He, Xe (enriched to 86\% \Xe), and N$_2$ in the various ratios shown in Table \ref{Table}. The cells are broadly divided into two categories containing high (50-100~Torr) and low (5-10~Torr) partial pressures of Xe.  Spheres were used because Eq.~\eqref{ShiftRatio} is strictly valid only for the case of a uniform spherical distribution of Rb magnetization for which the net average through-space dipole field is everywhere zero; effects due to imperfect geometry will alter the $^3$He shift only, because \KHe\ is on the order of unity, whereas \KXe\ is two orders of magnitude larger. The small ``pull-off" volume that results from cell fabrication (1-3\% of the total cell volume in our case) and an inhomogeneous laser intensity through the cell can both give rise to imperfect geometry. We note that geometrical shifts due to the nuclear magnetizations are less significant: rapid diffusion (compared to spin exchange) more readily guarantees a near-spherical distribution, and the \He\ magnetization is not inverted when the Rb magnetization is reversed.  In the case of \Xe, the magnetization is quickly attenutated/inverted because of rapid spin exchange; however, low $\left[\mathrm{Xe}\right]$ and low $\gamma_{\mathrm{Xe}}$ make any residual shift negligible. We have determined through a combination of numerical modeling and experimentation that geometrical shifts amounted to no more than a 1-2\% effect in even the most extreme cases.

\noindent

NMR free-induction decays (FIDs) were acquired at 67.6~MHz (\He) and 24.5~MHz (\Xe) in a horizontal-bore 2~T superconducting magnet (Oxford).  The Apollo (Tecmag) console is equipped with room-temperature shims and a gradient-coil set for imaging. The probe is a 35~mm diam Helmholtz coil immersed, along with the cell, in a safflower-oil bath contained in an Al-block reservoir with a plate-glass window to admit laser light. The probe could be tuned \textit{in situ} from one nucleus to the other by manually switching in/out additional capacitance without otherwise disturbing the apparatus. The Al block was heated with air that flows past an external filament heater. The IR-transparent oil bath reduced temperature inhomogeneity across the cell (observed with the laser on to be as large as 20\ $^{\circ}$C in a flowing-air oven) to $< 1\ ^{\circ}$C. A 30 W diode-laser array model A317B (QPC Lasers), externally tuned to the 795~nm $\mathrm{D}_1$ resonance and narrowed to $\approx0.3$~nm with a Littrow cavity \cite{Chann2000}, was mounted on an optical table with the optical axis aligned with the magnet bore (and the cell) for SEOP; the maximum narrowed output was $\approx20$~W. The quarter-wave plate in the optical train was mounted in such a way as to allow precise and reproducible manual rotation about the vertical axis by 180$^{\circ}$ in order to rapidly reverse the Rb magnetization---this is accomplished in a time on the order of the characteristic optical pumping rate (a few tens of microseconds) after the waveplate has been rotated into place; in practice the reversal takes $\approx 0.5$~s.

\noindent

Prior to data acquisition, SEOP was performed on the cell for a time $\gtrsim 1$~h, sufficient to build up polarization in both nuclear species. An auto-shimming procedure was performed with the laser blocked using the $^3$He frequency spectrum to narrow the resonance line to $\approx5$~Hz. After unblocking the laser and allowing a SEOP steady state to be established, two FIDs were acquired with an intervening Rb magnetization reversal. This basic procedure took $<$~1~s to perform, minimizing the effects of static field drift; it was then repeated $\approx 15$ times before switching to the other nucleus. Provided $\left|\left[\mathrm{Rb}\right]\langle S_z\rangle\right|$ remains constant throughout the measurement, the resulting frequency spectra yield \He\ and \Xe\ shifts suitable for use in Eq.~\eqref{ShiftRatio}, although the analysis has several subtleties.

\begin{figure}[tp!]
    \centering
        \includegraphics[scale=.143]{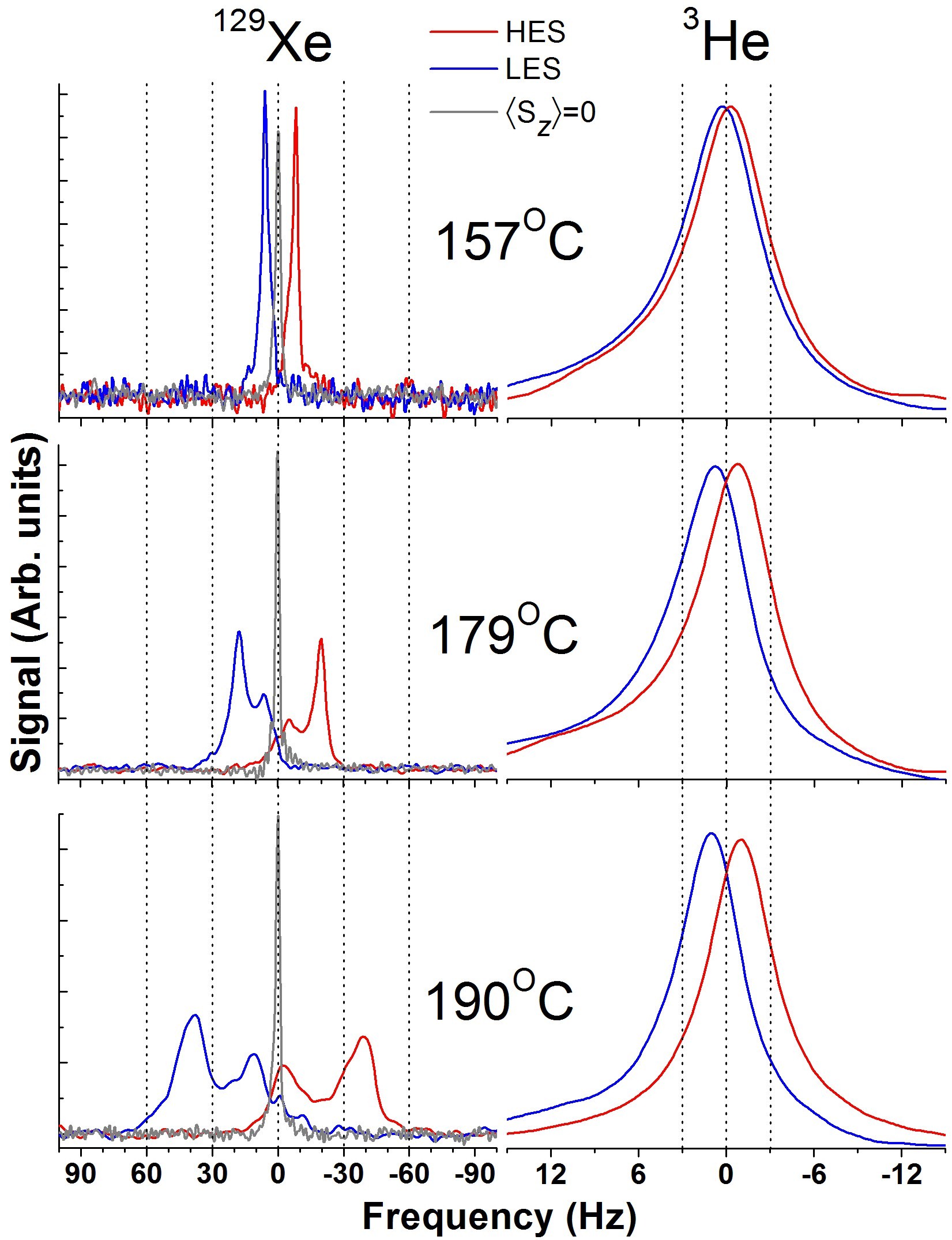}
   \caption{(Color online) Typical \He\ and \Xe\ spectra from cell 155B acquired one-after-the-other under steady-state SEOP conditions.  The narrow \Xe\ peak at 0~Hz was acquired with the laser blocked;  it has been amplitude-normalized to appear on the same graph.  The double peak, in the \Xe\ spectra at all but the lowest temperatures, represent regions of highly polarized and nearly unpolarized Rb vapor; the lines are broadened and begin to coalesce due to diffusion of \Xe\ between these two regions.  For \He, the much smaller frequency-shift dispersion and more rapid diffusion yeilds a single narrow peak in all cases.  The respective shifts in the spectral COM upon reversal of the Rb magnetization were used in Eq. \eqref{ShiftRatio} to extract \KXe.}
    \label{Data}
\end{figure}

\noindent

The \Xe\ spectra in Fig.~\ref{Data} have a characteristic two-peak structure for both the HES and LES compared to the narrow single peak acquired with the laser blocked (Rb unpolarized). In most cases, and particularly at higher temperatures, $\left[\mathrm{Rb}\right]\langle S_z\rangle$ is inhomogeneous due to lensing and attenuation of the laser light as it propagates through the cell; the spectrum is essentially a one-dimensional projection of the NMR frequency shift due to this inhomogeneous distribution of $\left[\mathrm{Rb}\right]\langle S_z\rangle$. However, the shape is also affected by diffusion. Diffusion of Xe is fast enough on the time scale of the FID that a given spin sees at least a partial average of frequency shifts. We hypothesize that the two-peak structure for $^{129}$Xe emerges as $\left[\mathrm{Rb}\right]$ increases due to the abrupt transition in the cell between fully polarized and nearly unpolarized Rb \cite{Walker1997}. We thus observed that there were substantial fractions of the cell volume at the highest temperatures where the Rb polarization was quite low. The situation is not unlike exchange between two chemically inequivalent sites, where the spectrum changes from two distinct peaks in the limit $\tau_{ex}\Delta\omega \gg 1$ to a single motionally narrowed peak in the opposite limit. Here, $\tau_{ex}$ is the exchange time between sites and is analogous to the diffusion time $\tau_{d}$ across the cell; and $\Delta\omega$ is the difference in frequency for the two sites, analogous to the frequency difference between \Xe\ in contact with polarized and unpolarized Rb. In our \Xe\ data it is often the case that $\tau_{d}\Delta\omega\gtrsim 1$; for \He, with faster diffusion and a much smaller frequency-shift dispersion, we have $\tau_{d}\Delta\omega\ll 1$, corresponding to a single-peak spectrum (see Fig.\ \ref{Data}).

\begin{table}
  \begin{tabular}{| l || c | c | }
    \hline
    Cell & Xe:N$_{2}$:He (Torr) & \KXe \\ \hline
    155A & 5:160:2200 & 495$\pm$6 \\ \hline
    155B & 10:250:2300 & 490$\pm$5 \\ \hline
    155C & 10:168:2300 & 530$\pm$9 \\ \hline
    150A & 50:175:1000 & $-$ \\ \hline
    150B & 110:350:2040 & $-$  \\ \hline
    155D & 50:172:1200 & $-$  \\ \hline
 \end{tabular}
 \caption{Summary of cell contents. All cells are sealed 7 mm i.d. uncoated Pyrex spheres.  Quoted pressures are referenced to 20~$^{\circ}$C and the Xe pressure is subject to $\approx50\%$ uncertainty due to the filling procedure.  \KXe\ is computed for each of the low-\XeD\ cells from the weighted average of that cell's data; we have excluded the high-\XeD\ cells because of their anomalous behavior at high $T$.}
 \label{Table}
\end{table}

\noindent

We note that the shapes of the HES and LES \Xe\ spectra in Fig.\ \ref{Data} are symmetric about the narrow spectrum acquired with unpolarized Rb. This was not always observed, although it is true of all data presented here. Our laser spectrum was narrow enough ($\approx 140$~GHz) that the optical pumping was affected by the $\approx 75$~GHz Zeeman shift of the D$_1$ resonance in a 2~T field for $\sigma+$ compared to $\sigma-$ light. For some data a small reproducible laser-tuning adjustment was made after flipping the quarter-wave plate, in order to keep the laser spectrum in the same position relative to  the absorption line.  The symmetry of the HES and LES spectra was used as an indicator that this adjustment had been made properly.  For other data we broadened the laser to $\approx$~950~GHz to guarantee symmetric absorption.

\noindent

The analysis of spectra like those in Fig.~\ref{Data} rests on the condition that the noble-gas magnetizations are uniform across the cell at all times, i.e., that the diffusion time across the cell $\tau_{d}$ is much shorter than the spin-exchange time for each species.  We measured the Xe diffusion coefficient using a standard pulsed-gradient technique \cite{Stejskal1864} to be $D_{\rm Xe} = 0.53 \pm 0.05$~cm$^2$/s in cell 150A at 170~$^{\circ}$C.  We thus have $\tau_{d}\approx d^2/6D_{\rm Xe}\approx 150$~ms. The measured value of the spin-exchange time for \Xe\ was at least 5~s in all cases. We acquired magnetic resonance images of cell 150A at 170 $^{\circ}$C that further verified the homogeneity of the nuclear magnetization. This condition, which is even more readily satisfied for \He, assures that all Rb spins in the cell are weighted equally in the NMR spectrum. This means that the shift in the spectral ``center of mass" (COM) that occurs when the Rb magnetization is flipped corresponds to the volume-averaged frequency shift for both species, regardless of which regime of diffusion-driven exchange (discussed above) holds for either species. The \Xe\ FIDs are multiplied by an apodizing exponential with a characteristic decay time about four times smaller than that of the FID. The subsequent fast Fourier transform produces a single broad symmetric spectral line that peaks at the spectral COM, the value of which is unchanged by the apodization procedure. As the raw \He\ spectra already consist of a single symmetric peak, they need no further analysis prior to measuring the shift. Thus, the shifts $\Delta\nu_{\rm He}$ and $\Delta\nu_{\rm Xe}$ are determined by comparing the respective HES and LES spectra and measuring the shift in a single peak. We note that for the equal-weighting assumption to be satisfied, the radio-frequency excitation field has to be homogeneous over the cell; if not, the shift ratio can depend weakly on the flip angle used (as we observed in our early data). We later employed a larger Helmholtz probe coil (sacrificing some sensitivity) to mitigate this problem.

\noindent

\begin{figure}[tp!]
    \centering
        \includegraphics[scale=.145]{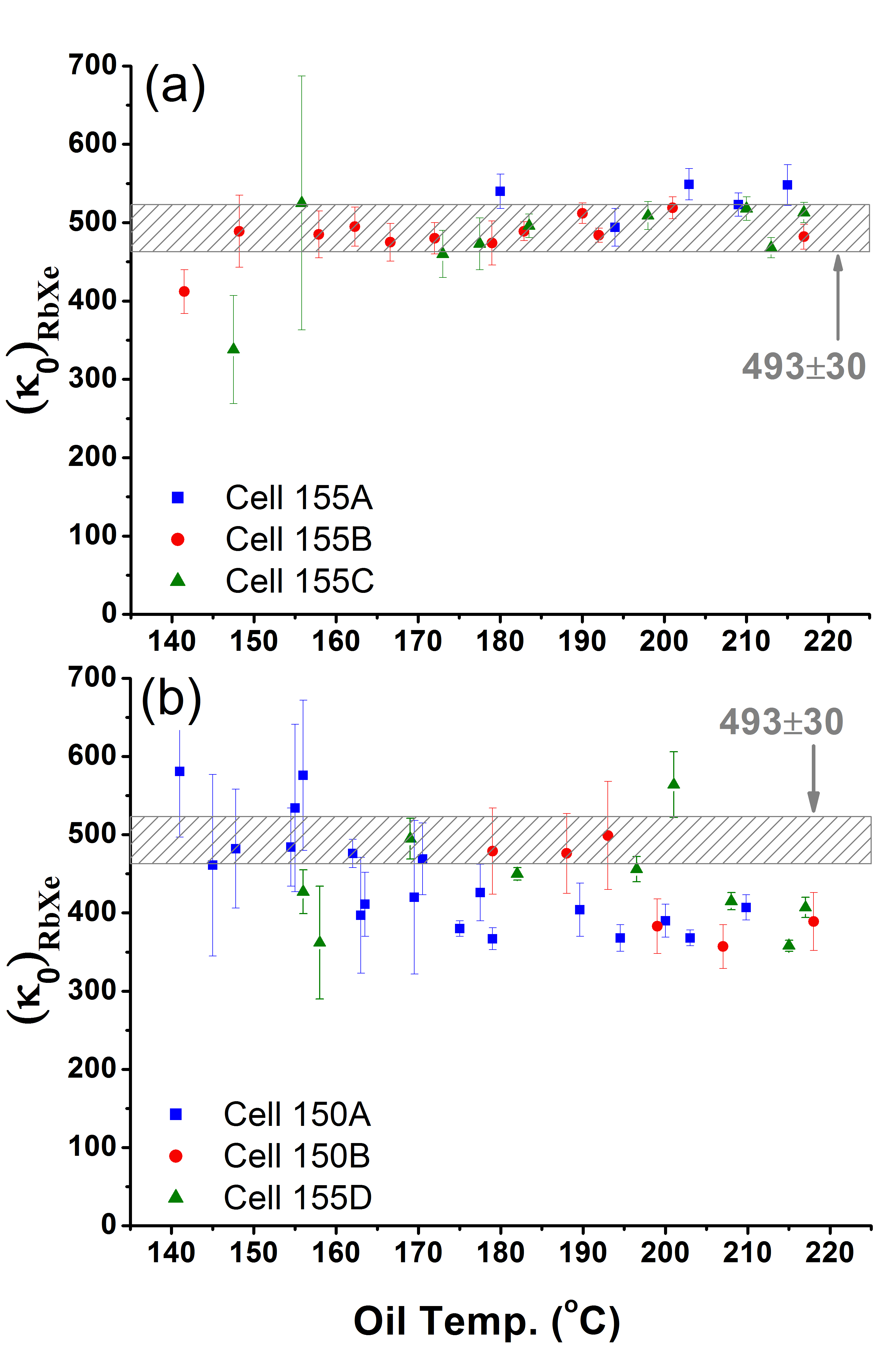}
    \caption{Enhancement factor \KXe\ plotted vs. temperature for (a) three low-\XeD\ cells and (b) three high-\XeD\ cells.  The weighted average of all the low-\XeD\ data points in (a) is 493, with the estimated uncertainty shown by the hatched region.  We take these temperature-independent data to represent the bet estimate of \KXe.  The identical hatched region is shown in (b) for comparison: the high-\XeD\ data are consistent with the low-\XeD\ data up to about 175~$^{\circ}$C; the $\approx 20$\% drop-off at the highest temperatures is not understood.}
    \label{KvsT}
\end{figure}

The calculated values of \KXe\ are plotted vs.\ temperature $T$ for the three low-[Xe] cells in Fig.~\ref{KvsT}a and for the three high-[Xe] cells in Fig.\ \ref{KvsT}b. The error bars shown reflect only the statistical uncertainty in the measured frequency-shift ratio and do not include the uncertainty in \KHe; they are dominated by the large relative uncertainty in the small \He\ frequency shifts. The low-[Xe] cells (both individually and collectively) show no significant temperature dependence between 140-220~$^{\circ}$C; the weighted average of all of these points yields an uncertainty of $< 1\%$. If we take the same weighted average on a cell-by-cell basis, there is a larger spread (see Table \ref{Table}), suggesting some unknown systematic errors at the few-percent level: these could include, for example, small cell-dependent geometrical effects. We accordingly increased the uncertainty in the shift ratio, which is represented by the hatched range in Fig.~\ref{KvsT}a. Finally, we add the 1.8\% uncertainty in the value of \KHe\ \cite{Romalis1998} in quadrature to this range to arrive at our final result in Eq.~\eqref{KappaValue}.

\noindent

Below $T\approx 175$~$^{\circ}$C, the data for the high-[Xe] cells are generally consistent with the hatched range (reproduced in Fig.~\ref{KvsT}b for comparison) that characterizes the low-[Xe] data. However, at the highest temperatures the measured shift ratio drops by about 20\%. These ten or so data points out of 60 acquired for all 6 cells are at the extremes of high temperature, high [Rb], and rapid Rb spin destruction (due to higher [Xe]); yet, we are unable to connect these physical conditions in a plausible way to the observed systematic depression of the shift ratio. We considered whether fast Rb-$^{129}$Xe spin exchange might lead to a violation of our fundamental assumption of uniform nuclear magnetization, but this would \textit{increase} the shift ratio by preferentially weighting the regions of higher Rb magnetization in the $^{129}$Xe spectrum. We also tested for extreme geometrical effects by remeasuring the shift ratio for both high- and low-[Xe] cells at a given temperature after significantly decreasing the laser power. The $^{129}$Xe spectrum changed dramatically under these conditions, but the shift ratio was unchanged within error. We note that this test also served as a check on the robustness of the COM data-analysis method.

\noindent

Schaefer \textit{et al.} \cite{Schaefer1989} calculated $\EqKXe=726$ and measured $\EqKXe=644\pm260$ by mapping the \Xe\ and $^{83}$Kr NMR spectra with the Rb EPR shift at low field.  The dominant source of error in their measurement came from modeling the Rb polarization. The theory that Schaefer \textit{et al.} present favors a zero to weakly-positive temperature dependence.  The anomalous behavior of our high-\XeD\ cells at high-T is neither consistent from cell to cell nor consistent with a plausible theoretical temperature dependence.  We thus take the global average of the low-[Xe] data as expressed in Eq.~\eqref{KappaValue} as our best estimate of \KXe, independent of temperature.  

\noindent

\indent The authors are grateful to T.~G.~Walker and G.~Laicher for helpful discussions. This work was
supported by the National Science Foundation (PHY-0855482).

\bibliography{References_Complete}

\end{document}